# The Noether charges of all analytic Lagrangians associated with a scale invariant action

Erik D. Fagerholm*, Robert Leech

Centre for Neuroimaging Sciences, Department of Neuroimaging, IOPPN, King's College London

* Corresponding author: erik.fagerholm@kcl.ac.uk

**Abstract**

Here we consider scale invariant dynamical systems within a classical particle description of Lagrangian mechanics. We begin by showing the condition under which a spatial and temporal scale transformation of such a system can lead to a symmetry by leaving the action unchanged. We then derive the form of all analytic Lagrangians that possess such a symmetry under change of scale. Finally, we write down the Noether charges of all analytic Lagrangians that are associated with a scale invariant action.

**Scale transformations:** Considering a Lagrangian for a driven dynamical system, i.e. one with explicit time-dependence: $\mathcal{L}(q, \dot{q}, t)$, we define the scale transformation of the position variable, $q$, as follows:

$$q' = \lambda q \qquad [1]$$

where $\lambda$ is a constant.

Similarly, we define the scale transformation of time, $t$, as follows:

$$t' = \lambda^\alpha t \qquad [2]$$

where $\alpha$ is a constant.

Using [1] and [2] we see that the velocity, $\dot{q}$, scales as:

$$\dot{q}' = \frac{q'}{t'} = \frac{\lambda}{\lambda^\alpha} \frac{q}{t} = \lambda^{1-\alpha} \dot{q} \qquad [3]$$

**Symmetries under change of scale:** The scaled action, $S'$, associated with a scaled Lagrangian $\mathcal{L}' = \mathcal{L}(q', \dot{q}', t')$ is defined via:

$$S' = \int \mathcal{L}' dt' \qquad [4]$$

And a Lagrangian is scale invariant if the following condition holds:

$$\mathcal{L}' = \lambda^n \mathcal{L} \qquad [5]$$

where $n$ is a constant.

We can therefore re-write [4] by using [5]:

$$S' = \lambda^n \int \mathcal{L} \left( \frac{dt'}{dt} \right) dt' \qquad [6]$$

and we can replace $\frac{dt'}{dt}$ in the Jacobian that accounts for the change of integration variable in [6] by using [2], such that:

$$S' = \lambda^{n+\alpha} \int \mathcal{L} \, dt = \lambda^{n+\alpha} S \qquad [7]$$

We see from [7] that there exists a symmetry ($S = S'$) under scale transformations in space and time if the Lagrangian [5] scales inversely with time [2], i.e. if:

$$n = -\alpha \qquad [8]$$

**The family of all scale-symmetric analytic Lagrangians:** We can write a generalised Lagrangian, $\mathcal{L}_g$, as a sum over power terms:

$$\mathcal{L}_g = \sum_{x,y,z} C_{x,y,z} q^x \dot{q}^y t^z \qquad [9]$$

where $C_{x,y,z}$ is an arbitrary expansion coefficient.

Using [1], [2] and [3] we see that [9] scales as follows:

$$\mathcal{L}_g' = \lambda^x \lambda^{(1-\alpha)y} \lambda^{\alpha z} \sum_{x,y,z} C_{x,y,z} q^x \dot{q}^y t^z \qquad [10]$$

Using [5] and [8] we see that if [10] is to be symmetric under scale then each term of [10] must scale with the same power $-\alpha$, such that:

$$\mathcal{L}_g' = \lambda^{-\alpha} \mathcal{L}_g \qquad [11]$$

We can then equate the exponents in [10] and [11], such that

$$x = (\alpha - 1)y - \alpha z - \alpha \qquad [12]$$

Using [12] we can re-write [10] as a summation over only two indices $y$ and $z$:

$$\mathcal{L}_g = q^{-\alpha} \sum_y C_y q^{y(\alpha-1)} \dot{q}^y \sum_z C_z q^{-z\alpha} t^z \qquad [13]$$

Equation [13] describes all analytic Lagrangians associated with a scale invariant action.

**The Noether charges of all scale-symmetric analytic Lagrangians:** According to Noether's theorem[1], the quantity $I$ (the Noether charge) is a constant of motion:

$$I = \left(\mathcal{L} - \dot{q}\frac{\partial \mathcal{L}}{\partial \dot{q}}\right)\delta t + \frac{\partial \mathcal{L}}{\partial \dot{q}} \delta q \qquad [14]$$

The spatial ($\delta q$) and temporal ($\delta t$) variations in [14] are defined as infinitesimal transformations. In order to use Noether's theorem we must therefore re-define the quantity $\lambda$ used so far to lie close to unity:

$$\lambda = 1 + \varepsilon \qquad [15]$$

where $\varepsilon$ is an arbitrarily small constant.

Using [15] we re-write [1]:

$$q' = \lambda q = (1 + \varepsilon)q = q + \varepsilon q \quad \rightarrow \quad \delta q = \varepsilon q \qquad [16]$$

Similarly, using [15] we re-write [2] and expand to first order:

$$t' = \lambda^\alpha t = (1+\varepsilon)^\alpha t \approx t + \alpha\varepsilon t \quad \rightarrow \quad \delta t = \varepsilon \alpha t \qquad [17]$$

We then re-write [14] using the definitions of $\delta q$ from [16] and $\delta t$ from [17] for infinitesimal changes under scale in space and time:

$$I = \left(\mathcal{L} - \dot{q}\frac{\partial \mathcal{L}}{\partial \dot{q}}\right)\alpha t + \frac{\partial \mathcal{L}}{\partial \dot{q}}q \qquad [18]$$

The expression in [18] is the Noether charge associated with systems that are symmetric under infinitesimal spatial and temporal scale transformations.

Using [13] and [18], we can then write down the expression for the Noether charges of all analytic Lagrangians associated with a scale invariant action:

$$I_g = \left((1-\alpha)\sum_y C_y y q^{y(\alpha-1)}\dot{q}^y \sum_z C_z q^{-z\alpha} t^z + \alpha \sum_y C_y q^{y(\alpha-1)}\dot{q}^y \sum_z C_z q^{-z\alpha} t^z\right) t q^{-\alpha} \qquad [19]$$

**References**

1    Noether, E. Invariante Variationsprobleme. *Nachrichten von der Königlichen Gesellschaft der Wissenschaften zu Göttingen. Mathematisch-physikalische Klasse*, 235-257 (1918).